# Time-resolved infrared photothermal imaging: From transient observations towards the steady-state

Dennis van de Lockand[1], Daan Wolters[1], Matz Liebel[1,*]

[1] *Department of Physics and Astronomy, Vrije Universiteit Amsterdam, De Boelelaan 1100, Amsterdam, 1081 HZ, The Netherlands*

* *email: m.liebel@vu.nl*

Mid-infrared photothermal microscopy is a highly promising imaging technique that enables spatially resolved vibrational fingerprinting. The combination of infrared induced heating with optical readout at visible wavelengths provides excellent spatial resolution while retaining the spectral observations of conventional infrared imaging. Most current implementations rely on long-duration illumination periods, to ensure sufficient heating and hence large signals. However, undesirable processes such as heat-diffusion degrade spatial resolution and the interplay between heat-induced refractive index changes and sample expansion adds additional uncertainties. Fundamentally, these issues stem from the difficulties associated with separating non-equilibrium and photoacoustic contributions from purely thermal signals. Highly time-resolved observations hold great promise for addressing these issues and are imperative for enabling future imaging modalities in this exciting temporal window. Here, we provide this much needed insight by employing widefield phototransient holography to phase-resolve optical responses from pico- to tens of nanoseconds following vibrational overtone excitation. We observe rapid transient-induced phase shifts, followed by heat-induced coherent expansion and thermalisation dynamics. Our observations provide a direct link between the photoacoustic and photothermal realms, thus not only offering much-needed insight for rationally optimising these exciting technologies but also avenues towards future all-optical stiffness and super-resolution imaging modalities.

**Background and introduction,**

Mid-infrared photothermal (MIP) microscopy enables label-free chemical imaging by combining infrared (IR) absorption spectroscopy with optical microscopy. At its core, MIP indirectly detects, and spatially resolves, IR-induced temperature changes through refractive index shifts and/or thermal expansion[1–9]. A variety of approaches exist for visualising these signals, including single-pixel detection schemes that capture thermal lensing[10] or thermal expansion[9]. Similarly, widefield methods such as phase-sensitive imaging and even intensity diffraction tomography have been employed[11,12]. Moreover, recent advances in the indirect detection of heat-induced signals via fluorescence intensity modulation offer promising new avenues for further improving detection limits down to single molecules[13–15]. Compared to alternative vibrational imaging modalities based on Raman scattering, MIP offers an exciting additional dimension to further tailor the experiments: time. This is due to the fact that MIP microscopy relies on IR-induced heating followed by a separate readout step[16] whereas Raman signals are generated within vibrational decoherence times of around 1 ps[17]. Accordingly, parameters such as the IR illumination duration, the IR excitation to readout delay, or the temporal width of the readout window can all be tuned to modulate the MIP signal[6,18–20].

In principle, processes such as intra-molecular vibrational energy redistribution, thermalisation with the surrounding media, thermal expansion and slow cooling should all be obtainable from adequately performed MIP measurements. The downside of this flexibility is that poorly adjusted parameters might have undesirable effects. Thermal diffusion-induced loss of spatial resolution, coupling between analyte size and thermalisation dynamics or the risk of accumulated sample heating all impact the measurement[21,22]. Several works have attempted to determine optimal measurement parameters with sub-nanosecond observation windows eliminating most obstacles[18,21,22]. However, a full picture based on systematic fundamental observations is missing. Pressing questions are related to the interplay of thermo-optically and thermal-expansion induced MIP-signals, or the possibility to further boost detection sensitivities by moving into the non-steady state regime[5]. Beyond addressing shortcomings, highly time-resolved MIP is likely to further expand the already exciting analytical capabilities. Chronological nanoscale insight into local thermalisation and expansion dynamics opens up a previously unexplored window into directly visualising material properties and concentration gradients at sub-diffraction-limited spatial scales[19,23].

Here, we provide this much needed advance by moving MIP microscopy from the nanosecond into the transient regime. More specifically, we perform time-resolved widefield phase imaging of vibrational overtone-excited samples on timescales ranging from pico- to several tens of nanoseconds. As such, we cover the thus-far unexplored window from non-steady state heating (hundreds of femto- to picoseconds) to intra-analyte thermalisation (many picoseconds), thermal expansion (pico- to nanoseconds) and the onset of thermalisation with the surroundings (nanoseconds). This extended temporal observation window links our first-of-a-kind observations with relevant fundamental phenomena as well as with the broader MIP literature in the nano- to microsecond range.

**Results,**

Figure 1 summarises our experimental approach which is loosely based on our recently introduced phototransient imaging schemes[24,25]. In brief, a widefield off-axis holographic transmission microscope records quantitative phase images using 1.5 ps probe-pulses centred at 515 nm, following established hologram-processing routes (Methods). To detect the IR-induced signals we perform pump-probe experiments where a tuneable sub-picosecond IR pulse photoexcites the sample of interest (Methods). The phase-difference between images recorded in the presence ($pump_{ON}$) or absence ($pump_{OFF}$) of IR-excitation directly yields the pump-induced differential phase images, $\Delta\varphi(t)$, as:

$$\Delta\varphi(t) = \arg\left(\frac{E_{ON}}{E_{OFF}}\right)$$

, where $E_{ON}$ ($E_{OFF}$) are the probe's holographically recorded electric fields in the presence (absence) of the IR-pump pulse. Recording differential images while systematically varying the pump-probe time-delay directly accesses the samples' temporal response from pico- to tens of nanoseconds and IR-pump wavelength tuning yields spectral fingerprints (Methods, Figure 1).

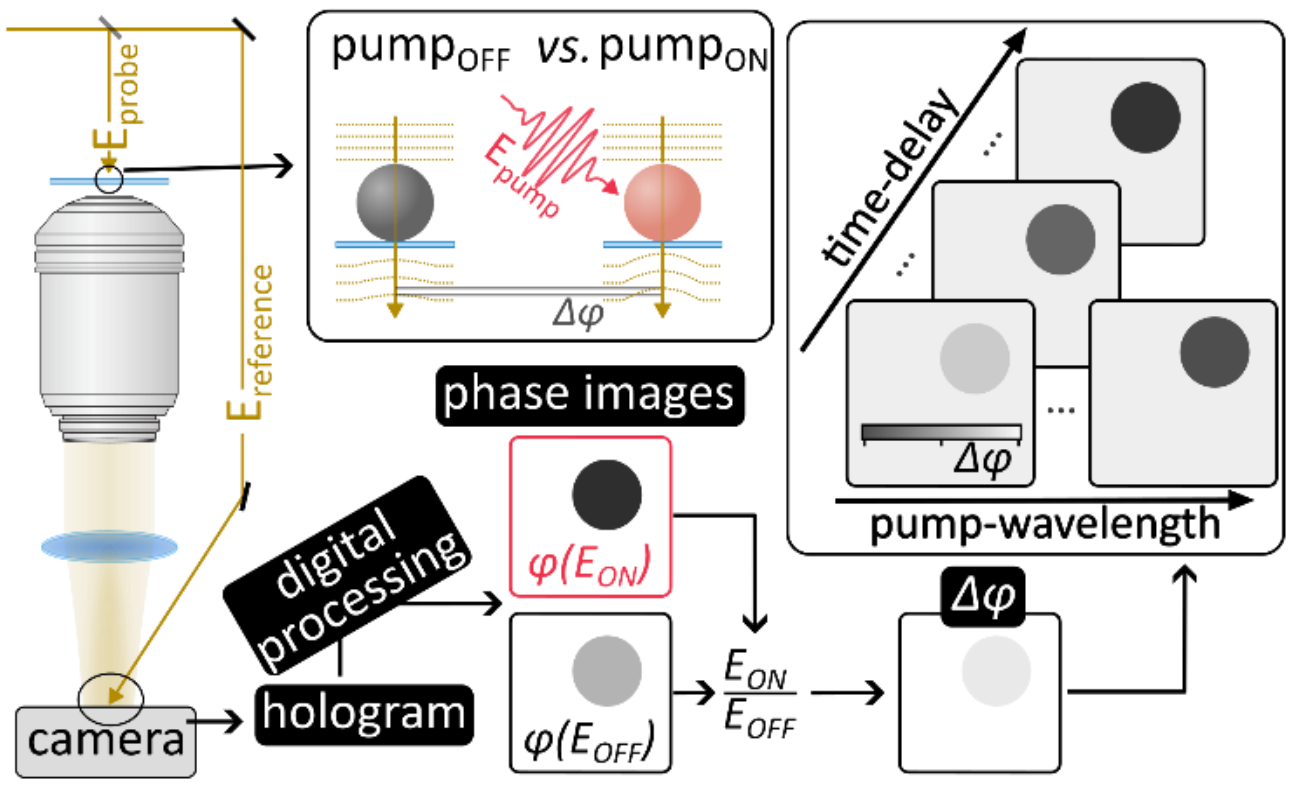

**Figure 1, Schematic of the principle and experimental implementation of IR-phototransient widefield imaging.** Holographic readout visualises phase changes due to IR overtone-excitation of dielectric samples which are extracted in a typical pump-probe differential imaging scheme. Systematically changing pump-wavelengths or pump-probe time-delays yields spatially resolved spectro-temporal information.

To benchmark our approach and showcase its broad applicability, we performed pump-probe experiments on 10 µm diameter polystyrene (PS) beads immobilised on glass in air (Methods). Figure 2a depicts a representative phase image of the sample recorded in the absence of the pump pulse. To uncover the temporal evolution of the system following vibrational excitation we recorded pump-probe delay dependent differential phase images while resonantly pumping the C-H stretching overtone transition of PS with a pump pulse centred at 1724 nm (Figure 2b).

No signal is present prior to time-zero, that is, when the sample is imaged before photoexcitation, indicating sufficient cooling between excitation events recurring at 500 Hz. Focusing on an individual PS bead, we encounter a coherent artifact around time-zero followed by a ~200 µrad signal that appears within the instrument response time (Figure 2b,c). Fascinatingly, a considerably larger oscillatory signal emerges within nanoseconds, with a peak-to-peak phase shift of around 4000 µrad. Importantly, pump-induced phase shifts observed on the glass substrate itself are negligible thus validating that the signals observed are originating from the PS beads.

To verify that the pump-induced signals are generated via vibrational overtone excitation and not through potentially competing nonlinear effects, we evaluated the signal's spectral and power dependencies (Figure 2d,e). Spectrally, we observed a qualitatively good agreement with literature values[26] (Figure 2d) and the signal scales linear with pump power (Figure 2e). Combined, these observations suggest that the PS-signals are, indeed, generated via one photon excitation of C-H stretching overtones at 5800 $cm^{-1}$.

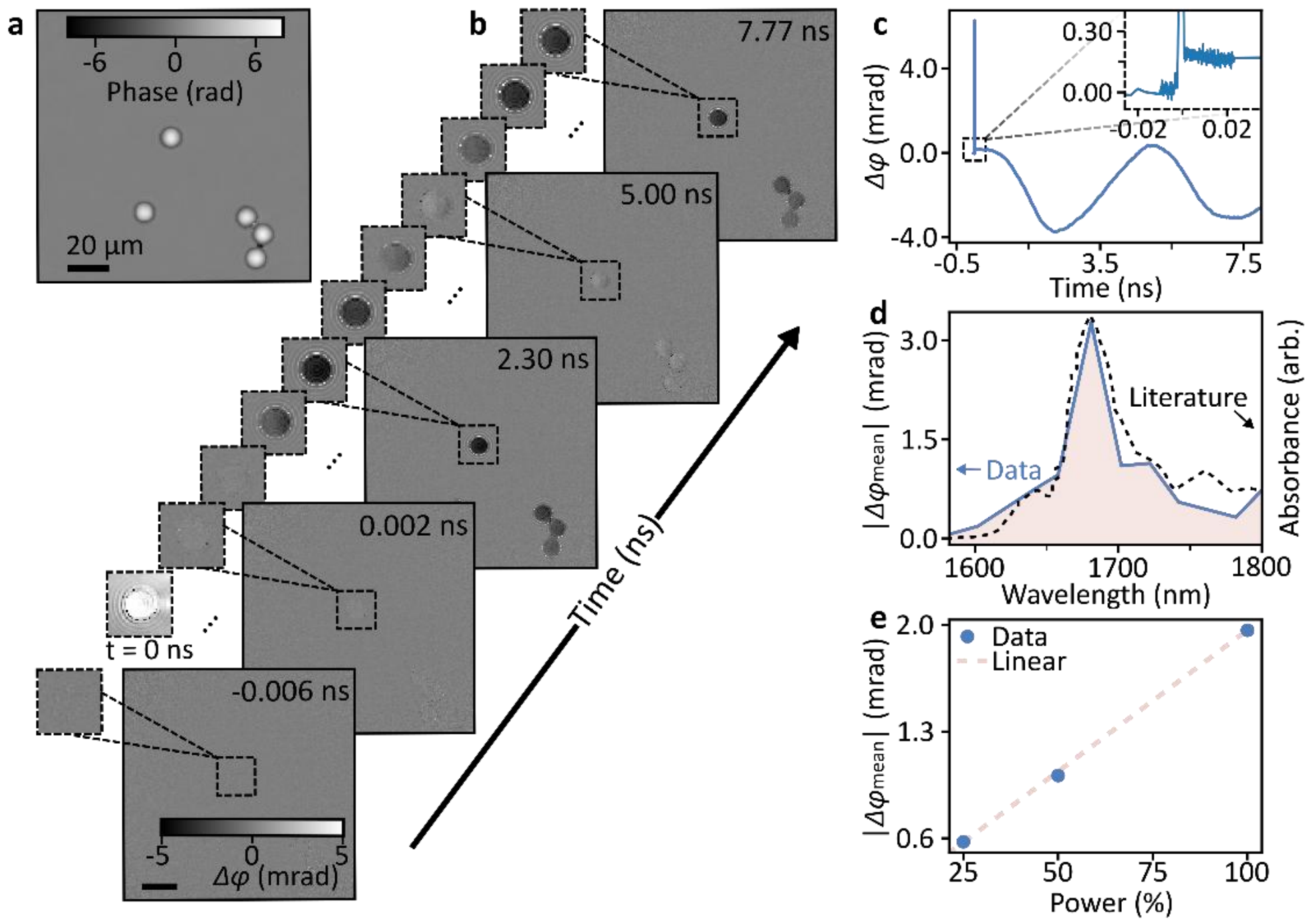

**Figure 2, IR pump-induced spectro-temporal dynamics of 10 µm polystyrene beads.** a) Phase image of the polystyrene (PS) sample showing five 10 µm beads on glass in air. The smaller than expected phase shift is due to phase-aliasing given the dramatic curvature of the sphere, compared to the diffraction limit. b) Differential phase images as a function of pump-probe delay using a pump fluence of roughly 56 mJ/cm$^2$ at 1724 nm and a 515 nm probe. The insets show an extended time-lapse of a single bead including its coherent artifact signal at time-zero. c) Mean differential signal of an individual bead. The sharp spike around time-zero is due to the coherent artifact. Inset: magnified signal around time zero with increased temporal sampling (Methods). d) Wavelength dependent differential signal for a single 10 µm PS bead obtained by temporally averaging one oscillation period between 0.27 and 5.6 ns (Supplementary Information 1) compared to a literature spectrum adopted from[26]. e) Pump power-dependence differential phase signal obtained as d).

To determine if and how the PS bead-size impacts the overall signal, we repeated the experiments described in Figure 2 for diameters of 10, 5 and 1 µm. Figure 3a shows representative time-dependent signals of individual PS beads as well as a region of roughly 150 aggregated 1 µm PS beads, to improve the signal-to-noise ratio of the 1 µm data. The corresponding frequency spectra (Figure 3b) indicate particle-size dependent acoustic signals, showing prominent frequencies of 0.22 GHz (10 µm), 0.43 GHz (5 µm) and ~1.9 GHz (1 µm) alongside some minor contributions at higher frequencies. The aggregated 1 µm beads exhibit a second frequency component around 2.1 GHz which could be due to size heterogeneity. These observations intuitively match the expected damped harmonic oscillator behaviour where an impulsively excited system resonates with a frequency that depends on its size and the material's speed of sound. The damping is governed by coupling to the surrounding media e.g. air with a small PS-glass contact region.

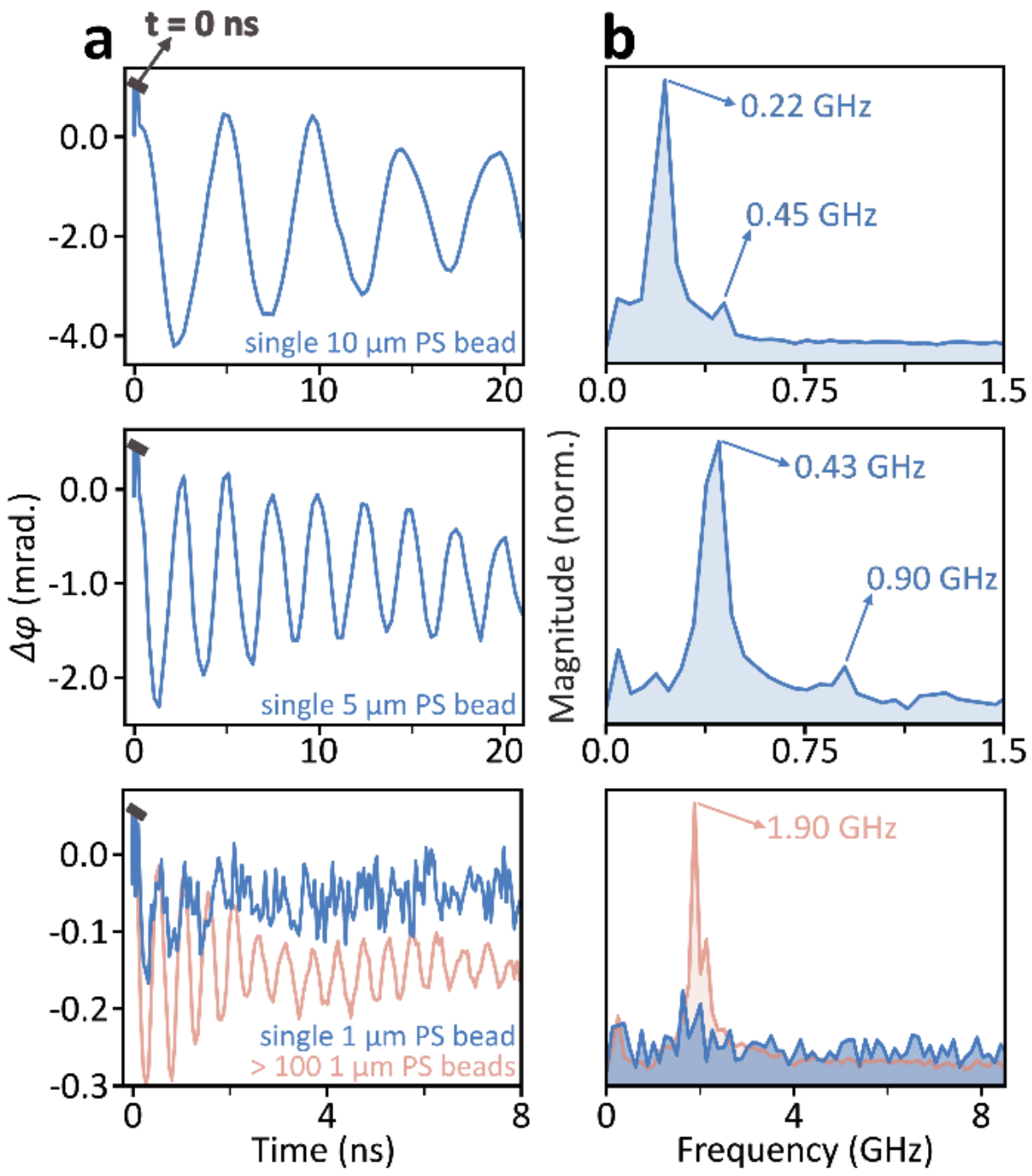

**Figure 3, Particle size-dependent coherent acoustic response.** a) Differential phase signals recorded for 10, 5 and 1 mum PS-beads on glass in air as a function of pump-probe delay. The coherent artifact around time-zero has been truncated as indicated by the black line. The slightly varying shapes of the oscillations are likely due to stepping errors of the very long mechanical delay line. b) Corresponding frequency spectra obtained via FFT. The temporal mean differential phase was subtracted prior to transforming.

Thus far we focused on PS-beads in air, but biologically and biomedically relevant microscopy is predominantly conducted in liquid or solid matrices[16,27,28]. Under these conditions, the analyte to matrix impedance matching is greatly enhanced. More specifically, PS exhibits a speed of sound of ~2370 m/s, water 1481 m/s and air 343 m/s[29–31]. In other words, oscillatory damping should be dramatically increased in water. To explore these aspects, we conducted experiments on 5 µm PS beads on glass in an aqueous environment. Our reliance on overtone excitation at 1724 nm conveniently circumvents the difficulties associated with strong mid-IR absorption when working at fundamental frequencies[32,33]. Figure 4a compares the signals obtained in the respective environments. Contrary to PS in air, we observe strong damping in water with the oscillation vanishing within less than one full period. These observations are in line with experiments performed on electronically excited gold nanoparticles[34–36]. A likely reason is that most of the thermal expansion induced pressure is efficiently coupled into the surrounding water. A closer inspection of the spatially resolved data (Figure 4b) reveals signatures of waves emerging from the particles, including interesting interference effects between them. Tracing the wavefront around the single 5 µm PS bead as a function of pump-probe delay allows estimating its velocity to being ~ 1438 m/s, close to the speed of sound in water[31] which suggests that we are observing an excitation-induced pressure wave (Supplementary Information 3).

## Discussion,

The combined observations warrant an immediate question: what is a MIP-suitable temporal measurement window that yields quantitative signals while ensuring diffraction limited resolution? Our observations suggest a complex interplay between sample geometry and temporal observation windows where coherently excited acoustic modes could randomly scale the signals, depending on their respective decoherence time. For poorly impedance matched samples, such as the air-PS system, extremely long pump-probe delays, potentially on the order of hundreds of nanoseconds might be necessary. An alternative would be to employ pump- or probe-pulse durations that make the observation of coherent acoustic modes impossible, through temporal convolution. Importantly, the situation described above is likely to also exhibit poor sample-to-substrate heat flow which renders heat-diffusion induced degradation of spatial resolution somewhat unlikely.

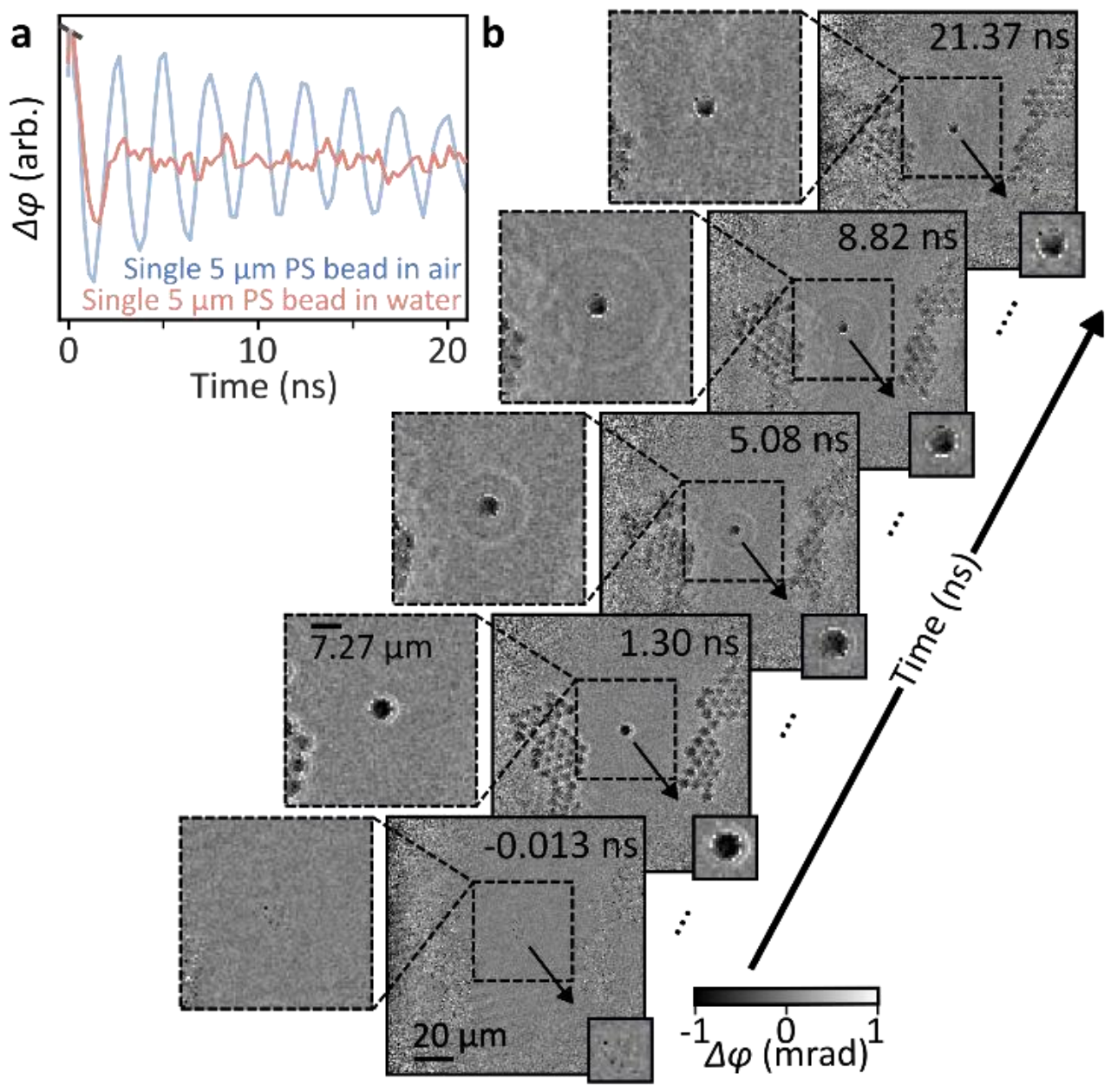

**Figure 4, Impedance matching governs acoustic responses.** a) Differential phase signals of a single 5 µm polystyrene bead in air versus water. The coherent artifact around time-zero has been truncated as indicated by the black line. b) Widefield differential phase images reveal the propagation of IR-pump induced pressure waves originating from all PS beads as enabled by the greatly enhanced acoustic PS-to-water coupling. Residual water background has been subtracted, raw data and subtracted backgrounds are shown in Supplementary Information 2.

From a sample perspective, the PS-water system is a more representative candidate where the analyte of interest and the matrix exhibit better-matched speeds of sound. This matching enables excitation-induced pressure waves to efficiently couple into the surroundings thus considerably dampening the coherent acoustic oscillations. In our case, a near-steady-state signal is obtained after roughly 2.5 ns for large, 5 µm diameter, PS beads (Figure 4). These observations are highly relevant to biological matter which exhibits speeds of sound comparable to that of PS[29,37]. In other words, our results suggest that biological particles in water with a diameter of 1 µm should reach steady-state in roughly 500 ps. This time-scale is comparable to optimised observation times which eliminate heat-diffusion induced loss of spatial resolution[18,22]. Leaving stress granules and biological condensates aside, this means that even widefield MIP microscopy should be able to interrogate essentially all biological systems in an artefact-free manner.

Beyond the direct implications for MIP microscopy our observations of a tiny, near-instantaneous, phase shift followed by an oscillatory signal that decays towards a much larger steady-state phase shift is surprising (Figure 2). MIP-signals are often attributed to originating from refractive index changes due to the thermo-optic coefficient and additional phase shifts governed by thermal expansion. Using literature values for the two ($dn/dT \approx -1.2*10^{-4}K^{-1}$, $\alpha \approx 7*10^{-5}K^{-1}$)[38,39] we can estimate the temperature induced phase-shift as:

$$\Delta\varphi(dT) = \frac{2\pi}{\lambda}(L\Delta n + n\Delta L) = \frac{2\pi}{\lambda}\left(L\left(\frac{dn}{dT}\right) + n\left(\frac{dL}{dT}\right)\right)$$

, where $n$ is the refractive index, $L$ is the length (height) of the sample and $\Delta L = \alpha L \Delta T$. Simple back-of-the-envelope calculations suggest that the absolute phase change due to the two contributions should be near-identical, differing by roughly a factor of 1.7. Temporally, a steady-state-temperature is reached within picoseconds, as governed by intra-molecular vibrational energy redistribution (Figure 5a). Thermal expansion is based on sound propagation and requires nanoseconds. The simplistic picture described above thus suggests that we should observe a near-instantaneous signal of a similar magnitude compared to the oscillatory response which is in stark contrast to our observations.

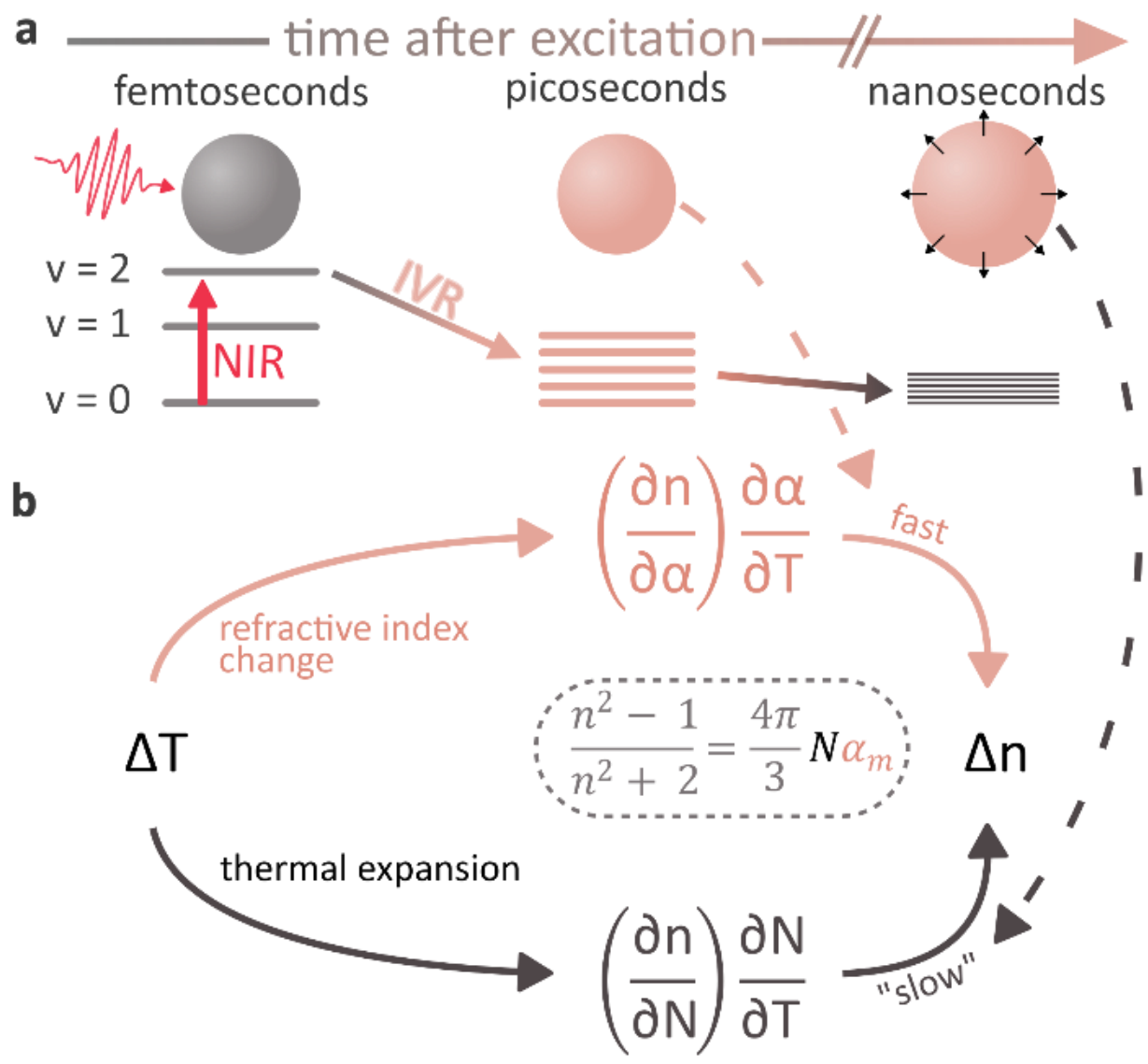

**Figure 5, Acousto-vibrational dynamics following vibrational overtone excitation.** a) Schematic of the cooling and expansion dynamics of a solid sphere following resonant excitation of a vibrational overtone. Intra-molecular vibrational relaxation (IVR) rapidly thermalises the non-steady-state excitation resulting in heating followed by lattice expansion. b) Schematic connecting a transient temperature change to the dynamics of refractive index changes and thermal expansion.

In order to rationalise our signals, it is instructive to consult the well-known Lorentz-Lorenz equation:

$$\frac{n^2 - 1}{n^2 + 2} = \frac{4\pi}{3} N\alpha_m$$

with $N$ being the number of molecules per unit volume and $\alpha_m$ the mean polarizability. This equation states that the change in refractive index is partly due to the change in mean polarizability of the molecules and partly due to the alteration in density because of the thermal expansion. In our macroscopic systems only the former happens on picosecond timescales. Albeit not explicitly indicating this dependency, the thermo-optic coefficient, $dn/dT$, combines both pure temperature dependencies as well as temperature-dependent density changes[39]. As such, highly time resolved observations decouple the respective contributions which allows explaining our observations. Figure 5b summarises the time-dependent contributions and their impact on the overall signal. Even though a steady-state temperature distribution is rapidly reached (Figure 5a), a large fraction of the thermo-optical signal change requires slow thermal expansion to allow the system to change its density. Translated to MIP microscopy, this model suggests that ultrafast approaches, such as our phototransient imaging scheme, are necessary to fully separate heat-induced from expansion-induced signals which is a distinct advantage for enabling quantitative observations for unknown analytes.

Given this increased information content, it should be possible to directly access relevant quantitates by modelling the temporal response. The poorly impedance-matched PS on glass in air system (Figure 3) is an ideal testbed for such approaches. While more complex finite element descriptions are likely necessary for arbitrary shapes, the system mentioned above can be approximated with spherically symmetric radial modes using Lamb's theory for breathing modes of an elastic sphere in a vacuum[41,42]:

$$F(\beta;\xi_i) = 4\frac{j_1(\xi_i)}{\beta^2\xi_i} - j_0(\xi_i) = 0$$

with $\beta = \frac{c_l}{c_t}$ being the ratio of the longitudinal, $c_l$, and transverse, $c_t$, sound velocities. $j_0(\xi_i)$ and $j_1(\xi_i)$ are the spherical Bessel functions required to solve for the breathing modes, where $\xi_i = f_i\pi d/c_l$, $f$ is the frequency and $d$ is the diameter[42]. To demonstrate how Lamb's theory can help rationalising our experimentally observed data (Figure 3) it is instructive to solve the equation outlined above to estimate breathing mode-dependent frequencies. Figure 6a shows solutions to this equation for PS using bulk sound velocities of polystyrene from literature ($c_l$ = 2370 m/s, $c_t$ = 1170 m/s)[29]. Based on these results, we compare estimated and measured frequencies (Figure 6b) and are, further, able to describe the time-domain oscillations using a simple model comprising the sum of two damped oscillators (Figure 6c). For both 5 and 10 µm PS sphere, we observe a close match between estimated and experimentally measured frequencies for both breathing modes with a minor discrepancy, around 3%, for the 5 µm case. Changing the diameter by 2.8%, to 5.14 µm, in line with typical size distributions for monodisperse microspheres, removes this discrepancy, making size-heterogeneity a likely explanation (Supplementary information 4). These observations suggest that the acoustic velocities of these macroscopic PS spheres closely resemble that of bulk polystyrene.

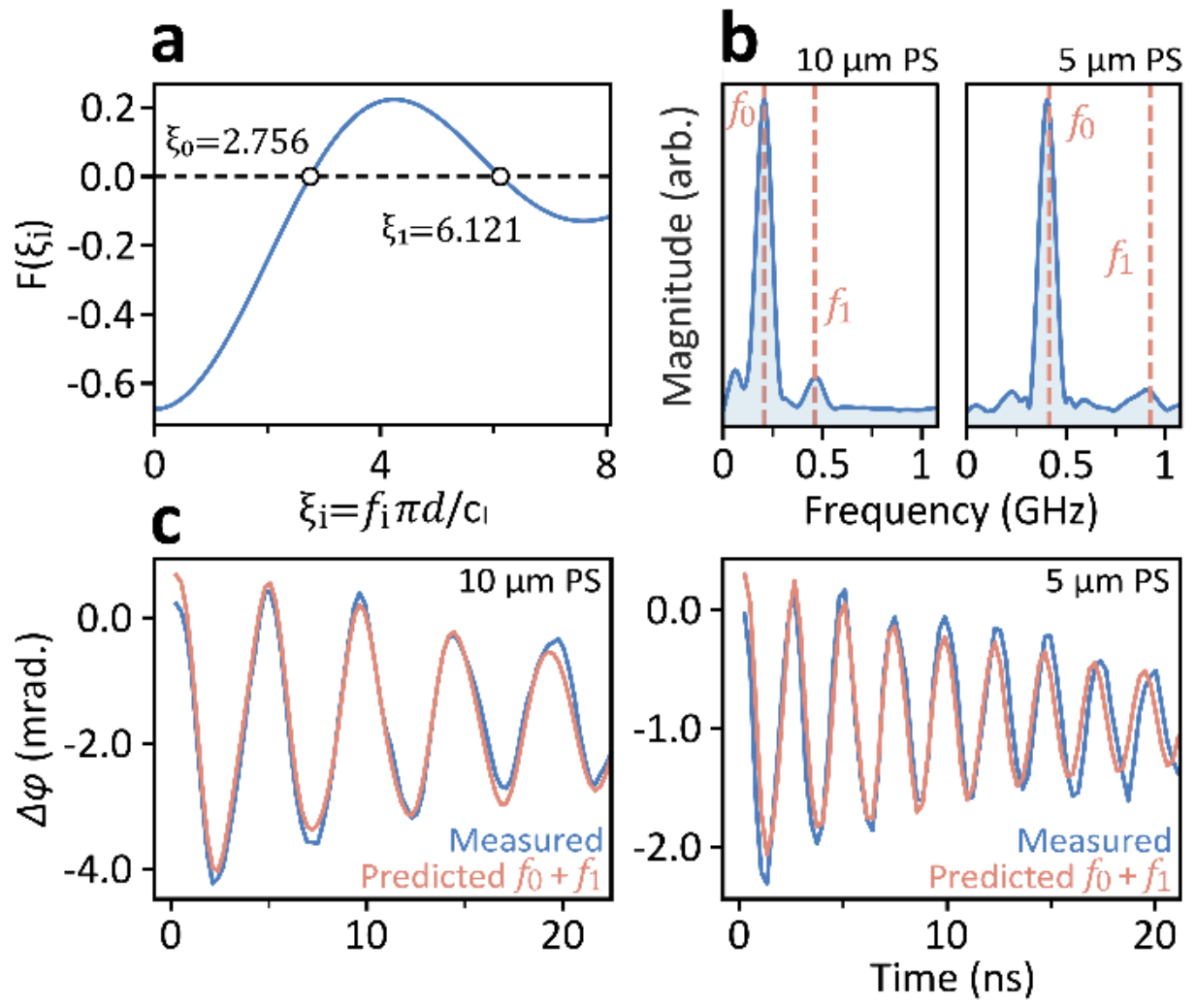

**Figure 6, Lamb's theory of a breathing sphere compared to experimentally measured frequencies.** a) The stress-free boundary equation for the breathing modes an elastic sphere $F(\beta;\ \xi_i)$ at a fixed $\beta$ determined from the literature longitudinal and transverse sound velocities of bulk polystyrene. b) Measured frequency spectra (blue) vs predicted frequencies of the breathing modes of the PS spheres (dashed vertical lines). Predicted frequencies are obtained from the $\xi$ *values in* a) combined with the nominal diameters of the 5 and 10 µm PS spheres, where $f_{0,\,10\mu m} = 0.2079$ GHz, $f_{1,\,10\mu m} = 0.4618$ GHz, $f_{0,\,5\mu m} = 0.4158$ GHz and $f_{1,\,5\mu m} = 0.9235$ GHz. The centre frequencies of the measured spectra, obtained via polynomial baseline detrending, Hanning and zero-padding the original measured trace (Supplementary information 4), are $f_{0,\,10\mu m} = 0.2059$ GHz, $f_{1,\,10\mu m} = 0.4655$ GHz, $f_{0,\,5\mu m} = 0.4045$ GHz and $f_{1,\,5\mu m} = 0.9052$ GHz. c) Measured oscillatory traces for the different sized PS spheres fit using the sum of two simple damped oscillators, fixing the predicted frequencies in b).

The procedure outlined above, unsurprisingly, reproduces the experimental observations given that both the particle sizes as well as their composition is known. While particle-sizes can be inferred experimentally, we require knowledge of literature-values for longitudinal and transverse speeds of sound of PS. In contrast, the direct observation of multiple breathing modes enabled by our time-domain measurements allows composition and literature-value independent parameter extraction by solving the equation outlined above, thus giving direct access to not only the longitudinal sound velocity but also the transverse component and consequently also stiffness moduli. To recap, each breathing modes' eigenfrequency can be described as:

$$f_i = \frac{\xi_i c_l}{\pi d}$$

Where $\xi_i$ is solely a function of $\beta = \frac{c_l}{c_t}$ for the assumed stress-free boundary conditions with $F(\beta; \xi_i) = 0$. In other words, the observation of two frequencies, combined with knowledge of the particle size, allows numerically determining the two unknown variables of interest, $c_l$ and $c_t$. Once obtained, Poisson's ratio can be determined from $\beta$ via $\upsilon = (\beta^2 - 2)/(2\beta^2 - 2)$. Additionally, it is possible to infer the longitudinal modulus $M = \rho c_l^2$, shear modulus $G = \rho c_t^2$, bulk modulus $K = M - 4G/3$ and Young's modulus $E = 2G(1 + \upsilon)$ if a density ($\rho$) estimate can be obtained.

| **Property** | **5 µm PS** | **10 µm PS** | **Literature PS[29]** |
|---|---|---|---|
| Longitudinal sound velocity (m/s) | 2326 ± 93 | 2397 ± 96 | 2370 ± 20 |
| Transverse sound velocity (m/s) | 1177 ± 47 | 1250 ± 50 | 1170 ± 10 |
| Poisson's ratio | 0.33 | 0.31 | 0.34 |
| Longitudinal modulus (GPa) | 5.6 ± 0.45 | 6.0 ± 0.48 | 5.8 |
| Shear modulus (GPa) | 1.4 ± 0.11 | 1.6 ± 0.13 | 1.4 |
| Bulk modulus (GPa) | 3.7 ± 0.30 | 3.8 ± 0.30 | 3.9 |
| Young's modulus (GPa) | 3.8 ± 0.30 | 4.3 ± 0.34 | 3.8 |

**Table 1,** Measured sound velocity and elastic properties of the PS spheres compared against literature values, stiffness values are calculated using a PS density of 1.04 g/cm$^3$. Uncertainties derived from 4% uncertainty of the PS spheres nominal diameters.

The measured acoustic and elastic properties of the PS spheres are summarized in Table 1 and compared with literature values for bulk PS obtained using Brillouin scattering[29]The calculated longitudinal and transverse sound velocities for both the 5 and 10 µm PS spheres agree well with its literature values of 2370 and 1170 m/s for bulk PS. Consequently, the corresponding Poisson ratio's and elastic moduli, are also extremely similar showing a close similarity between the PS spheres elastic behaviour and that of bulk PS.

**Summary and conclusion,**

To summarise, we widefield-visualised the phototransient response of dielectric materials to ultrafast vibrational excitation. Relying on vibrational overtones, rather than fundamental modes, allowed us to conduct near background-free experiments in aqueous solutions which directly translate to mid-IR based schemes. We observed instantaneous refractive index modifications due to polarizability changes followed by sample and substrate dependent thermo-acoustic responses on pico- to nanosecond timescales. Enabled by our capabilities to directly access these signatures through phase-based observations from pico- to several tens of nanoseconds we were able to devise solid-recommendations for an informed optimisation of MIP observations. Using the well-defined PS sphere on glass in air system, we were able to showcase that combing imaging-accessible parameters, such as the structure's size, with temporal-sampling of the oscillatory response of the system allows retrieving elastic parameters without the need for direct mechanical contact or knowledge of the refractive index[40,41].

Our first-of-a-kind observations offer exciting avenues for future work. Label-free time-resolved thermal expansion imaging accesses speeds of sound which can be translated into sample-stiffness. Similarly, decoherence timescales should, in principle, access sample composition beyond the diffraction limit where a porous or partially hydrated sample, such as a protein condensate, is likely to exhibit distinct signatures when compared to a solid sphere: information that is difficult to obtain with existing techniques. While our first steps towards full elastic sample characterisation use a spherical system that can be described by Lamb-theory we believe that future work using rigorous finite element modelling combined with spatial information and additional insight that provides density-estimates will allow extending the framework introduced towards full elasticity characterisation of unknown samples. As an example, there are biological specimens where careful experimental procedures allow density to be inferred from optical diffraction tomography based refractive index mapping[42]. Finally, the direct observation of pressure waves in aqueous environments offers exciting avenues towards ultra-resolution imaging. At acoustic frequencies, the wavelength and hence the spatial resolution limit is determined by the speed of sound, not the speed of light. Given the approximate five orders-of-magnitude difference between the two, high-frequency acoustic oscillators, such as small gold nanoparticles, should thus allow approaching nanometre-resolution label free imaging based on the experimental methodology described here. Moving forward, we anticipate biomedical and biologically relevant high-resolution applications of this approach. This transition will be facilitated by moving towards direct vibrational excitation in the spirit of MIP thus capitalising on approximately 100-times fold larger absorption cross sections.

**Methods,**

**Light generation,** Spectrally tuneable IR pump pulses, as characterised by measuring their frequency doubled spectra (*AFBR-S20M2VN, Broadcom*) are generated by a home-built multi-stage (non-collinear) optical parametric amplifier ((N)OPA), pumped by the second harmonic (515 nm) of an amplified Ytterbium:KGW laser (*Pharos PH2-20W-SP*, *Lightconversion* operated at 1 kHz, 2 mJ, 400 fs). In brief, <1 µJ of 1030 nm generate white light in a 10 mm YAG which is then amplified in a double pass NOPA tuned to broadband phase matching yielding 2-3 µJ of 150 $cm^{-1}$ broad tuneable light in the 650-980 nm region. A collinear OPA booster stage consisting of three BBOs[43] pumped by 500 µJ of 515 nm light then further amplifies this seed to obtain the tuneable IR-pump in the 1550-1850 nm range. Frequency doubling a small, picked off, fraction of the fundamental yields the 515 nm probe pulse.

**Phototransient microscopy,** A double-pass grating stretcher chirps the 515 nm observation light to roughly 1.5 ps following probe and reference wave generation by a 70:30 beamsplitter. A F = 250 mm lens focuses the probe into the sample plane of a home-build transmission microscope (20x, NA = 0.45 objective, *Motic*). An imaging lens, combined with a relay imaging/magnification system, then images the sample plane onto a CMOS camera (frame rate 1 kHz, HZ-21000-G, *Emergent Vision Technologies Inc.*) at a nominal magnification of 160x resulting an approximate field-of-view of 111.6x111.6 µm.

To perform off-axis holography, we interfere the signal wave with a reference wave on the same camera sensor where a diffraction grating (*Holo/Or*), placed conjugate with the camera via a 4F-system, generates the reference as one of its first diffraction orders. The time-delay between the probe and reference waves is carefully matched via a manual translation stage.

The IR pump is focused into the sample plane to an approximate spot size of 71x32 µm at full-width-half-maximum. A mechanical chopper (*MC2000B*, *Thorlabs*) modulates the pump at 500 Hz. Images are recorded in a single-shot manner, in the presence or absence of the pump pulse, at 1 kHz. Pump-probe time-delay dependent images are recorded using two computer-controlled translation stages. One for fine-stepping (M230.25, *Physik Instrumente (PI) SE & Co.KG*) and one for long distance delays (*FPB45* (2000 mm), *FUYU Automation*). The *FUYU* stage is operated in a double-pass configuration and controlled by an in house written, *Nexys A7-100T* based, controller. Active beam-stabilisation ensures beam displacement-free acquisition, even for >7 m mechanical delays.

**Holographic data processing and signal processing,** Hologram pre-processing is performed in an online fashion via *GPUDirect* using a *Mellanox® ConnectX-6 DX Dual Port 100GbE* network interface card and an *A4000 GPU* (*Nvidia*). In brief, images are streamed into the GPU where they are Fourier transformed. The momentum-separated k-space of the interference term is then extracted followed by an inverse Fourier transformation, using standard off-axis holography processing schemes[25]. The resulting complex images are then sent to the CPU and saved for further analysis. Differential phase images are obtained by dividing $E_{ON}/E_{OFF}$ fields on an image-by-image basis, for a 1000 images, followed by phase extraction. After phase extraction, the 1000 differential phase images are averaged giving a single averaged image for each pump-probe delay. Absolute phase shifts or linear x-/y-phase-gradients due to minor beam motion between images are removed by linearly fitting the phase background. The differential phase signals of individual beads, as shown in Figures 2c-e, 3a and 4a, are obtained by computing the mean phase over a square area fitting inside the spatial extend of the bead. The residual background signal originating from water (Figure 4) was removed using a Savitzky-Golay filter and a third order polynomial (see Supplementary Information 2 for processing steps and unprocessed data).

**Acquisition settings,** The fine stepping stage is used for all pre time-zero measurements as well as the inset shown in Figure 2c, all other delays are controlled via the *FUYU* stage. All experiments, besides the spectrally resolved data (Figure 2d) are performed using a 1724 nm IR-pump with an approximate duration of 200 fs. Pump fluences at the sample are 56-168 mJ/cm$^2$, besides the 1 µm PS sample that was illuminated at 670-730 mJ/cm$^2$. Due to the large measurement uncertainty associated with our power meter (*S425C-L*, *Thorlabs*), we performed the power-dependence reported in Figure 2e using pre-calibrated neutral density filters and a representative 100% pump fluence of roughly 56 mJ/cm$^2$. The powers used in Figure 2d were measured using an OD3 neutral density filter combined with an photodiode (*S122C*, *Thorlabs*).

**Sample preparation.** #1.5H coverglass was cleaned by 10 min sonification in acetone followed by 10 min in milli-Q water. The glass was then dried under a stream of $N_2$-gas. The polystyrene beads were diluted in absolute ethanol, pipetted and spread out onto substrate and left to dry. For the water measurements a second cover glass was taped onto the PS-containing glass using ~ 200 µm double sided sticky tape followed by filling the resulting gap with milli-Q water.

**Acknowledgement and Funding,**

We would like to thank Prof. Michel Orrit for insightful discussions and Prof. Freek Ariese for providing the PS samples. M.L. acknowledges funding by the European Union (ERC, PIRO, grant number: 101076859). Views and opinions expressed are however those of the author(s) only and do not necessarily reflect those of the European Union or the European Research Council. Neither the European Union nor the granting authority can be held responsible for them.

**References**,

## Supplementary Information:

# Time-resolved infrared photothermal imaging: From transient observations towards the steady-state

Dennis van de Lockand[1], Daan Wolters[1], Matz Liebel[1,*]

[1] *Department of Physics and Astronomy, Vrije Universiteit Amsterdam, De Boelelaan 1100, Amsterdam, 1081 HZ, The Netherlands*

* *email: m.liebel@vu.nl*

### Supplementary Information 1, Spectrally resolved and power-dependent measurements,

To provide representative spectral measurements, albeit the strongly time-delay-dependent signals (Figure 2c, main manuscript), we recorded temporal observation windows covering approximately half a period, as shown in Supplementary Figure 1a. The mean signals obtained from these measurements are shown in Figure 2d in the main manuscript. Supplementary Figure 1b show the second harmonics of the corresponding IR-pump spectra which were used to determine the excitation frequency.

Supplementary Figure 1c shows the power-dependent oscillatory response of the 10 µm PS beads as a function of pump-probe delay as used for determining the power dependence shown in Figure 2e of the main manuscript.

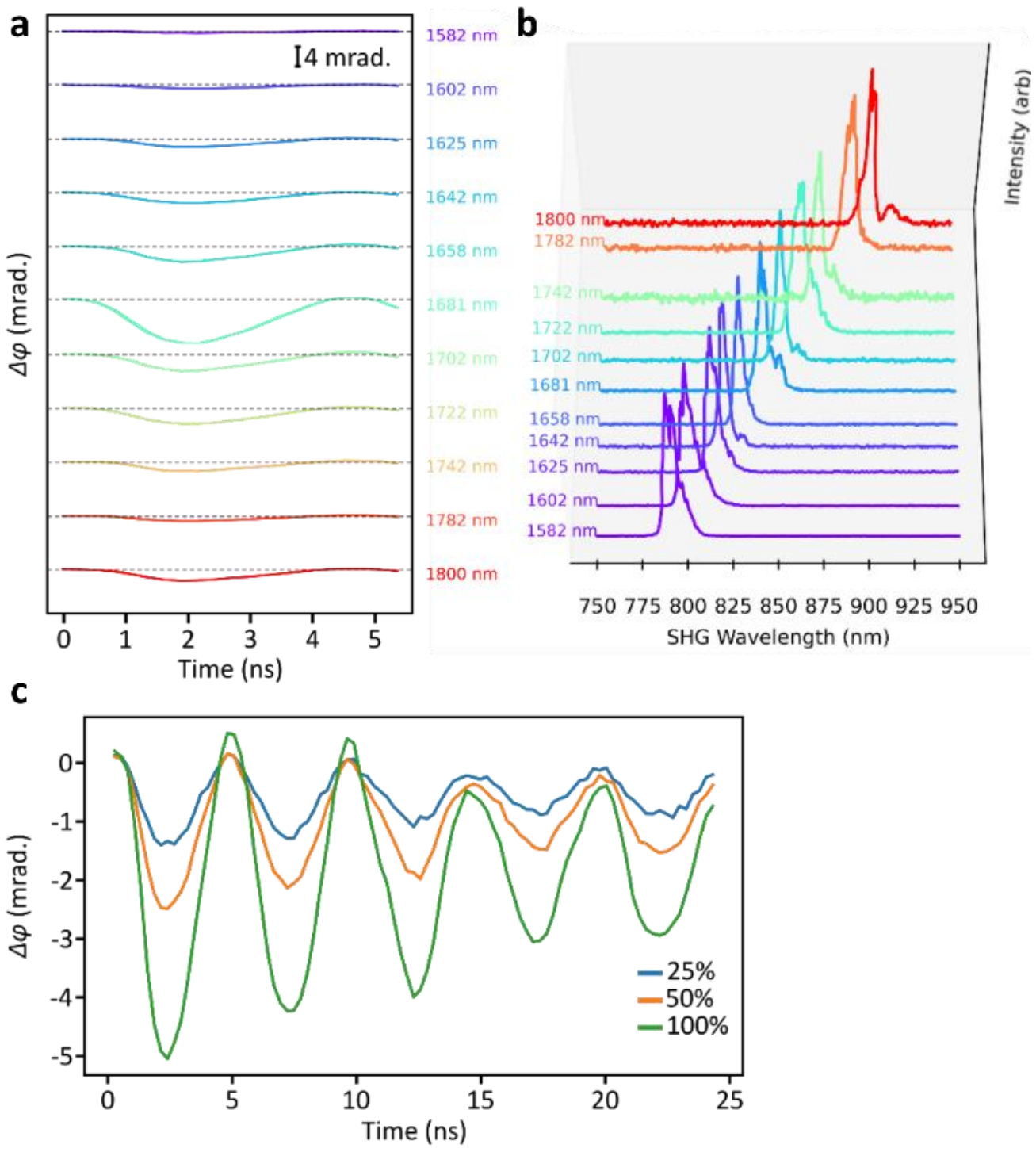

**Supplementary Figure 1, Additional signal characterisation.** a) Wavelength-dependent response of 10 µm bead sampled over approximately half a vibrational period. b) Corresponding second-harmonic spectra of the IR-pump employed. c) Sample response for varying relative excitation powers at 1724 nm.

**Supplementary Information 2, Residual water-background removal,**

Supplementary Figures 2,3 show the as-acquired and background-removed differential phase images for the 5 µm PS bead-measurements in water as shown in Figure 4 of the main manuscript. The corresponding background is shown in Supplementary Figure 4. The background of the 200x200 px phase image was estimated using a Savitsky-Golay filter with a window length of 115, and polynomial order of 3.

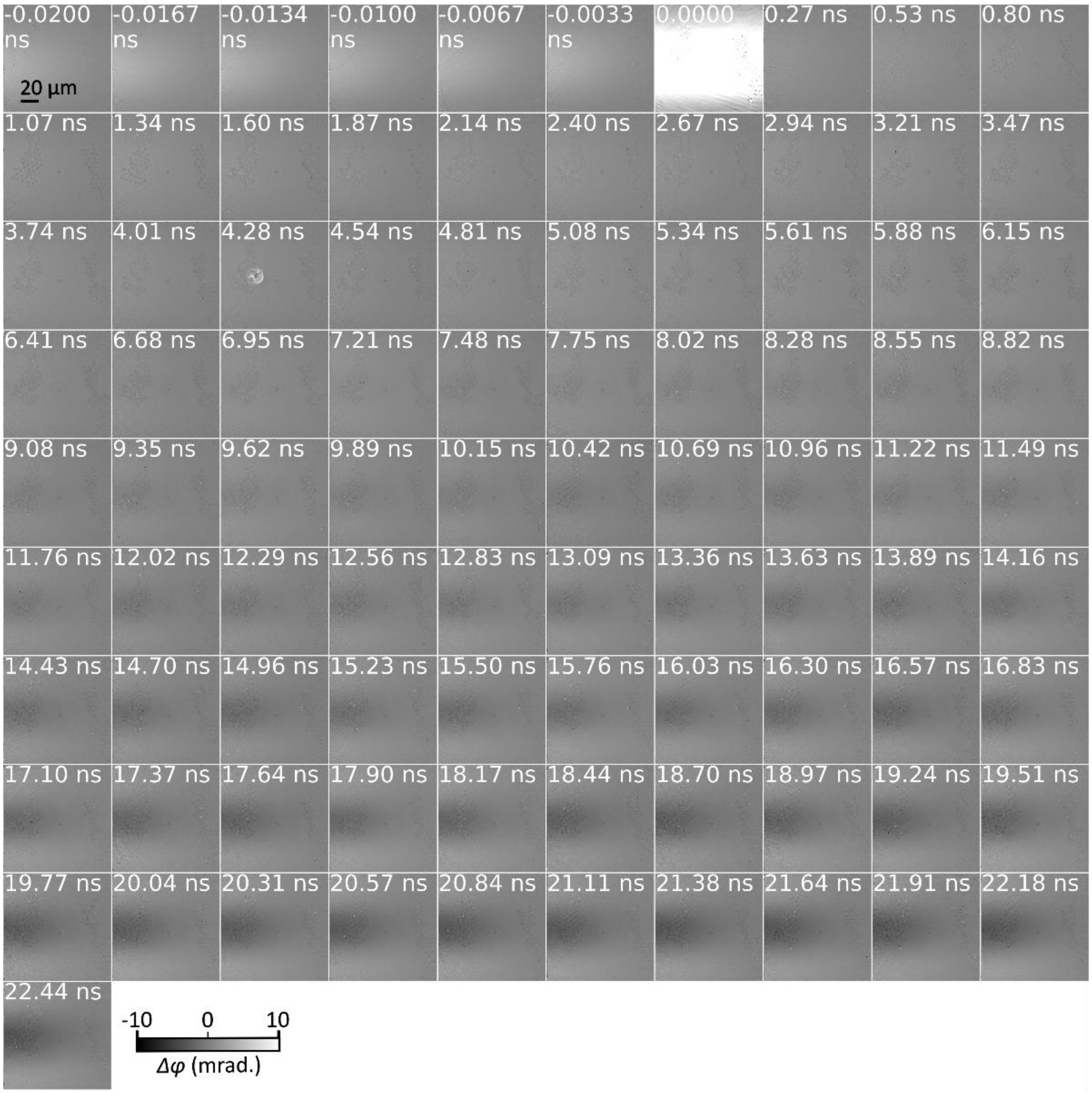

**Supplementary Figure 2,** As-acquired differential phase images of the 5 µm PS beads in water experiments discussed in Figure 4 of the manuscript.

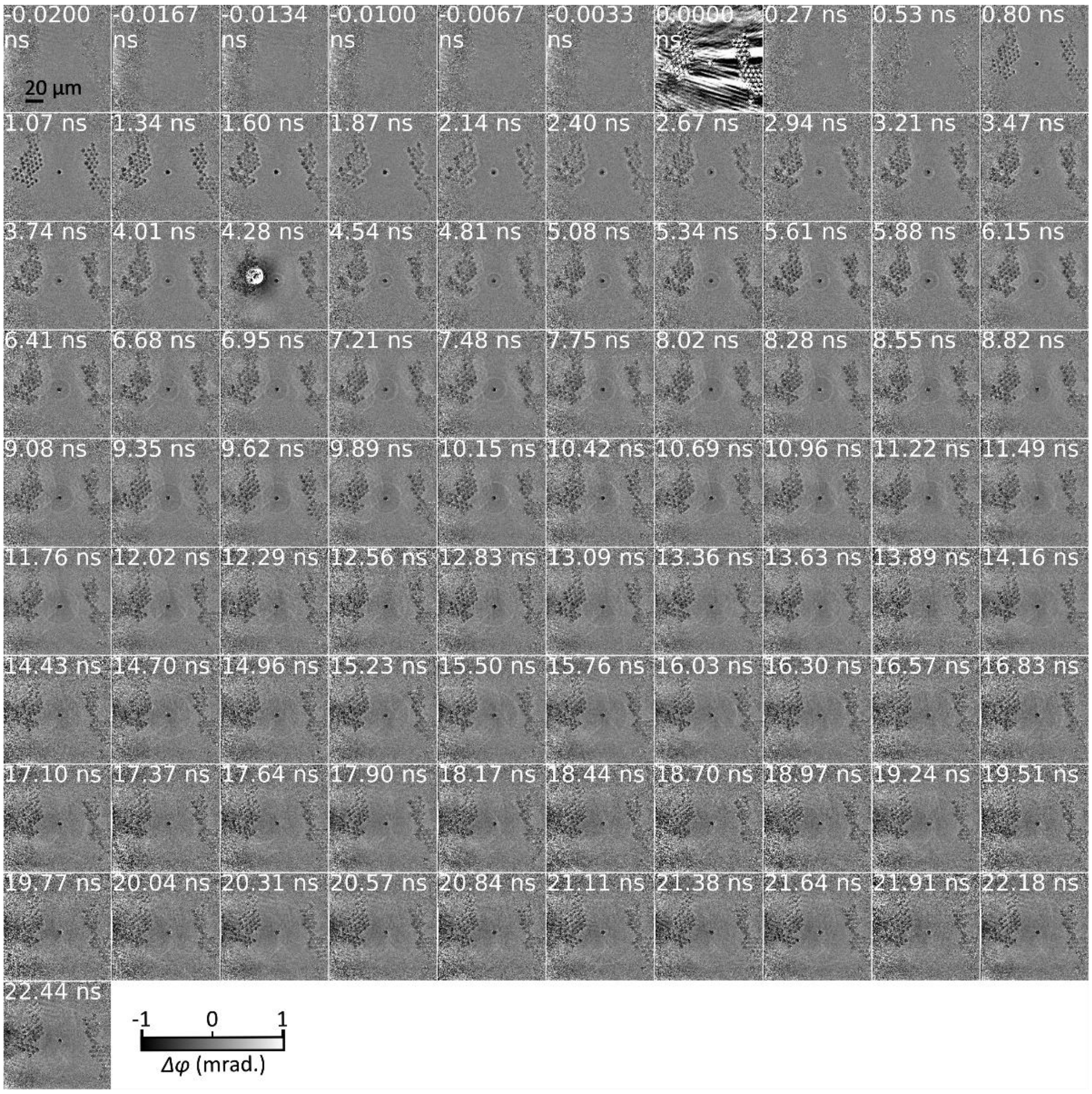

**Supplementary Figure 3,** Background-subtracted differential phase images of the 5 µm PS beads in water experiments discussed in Figure 4 of the manuscript.

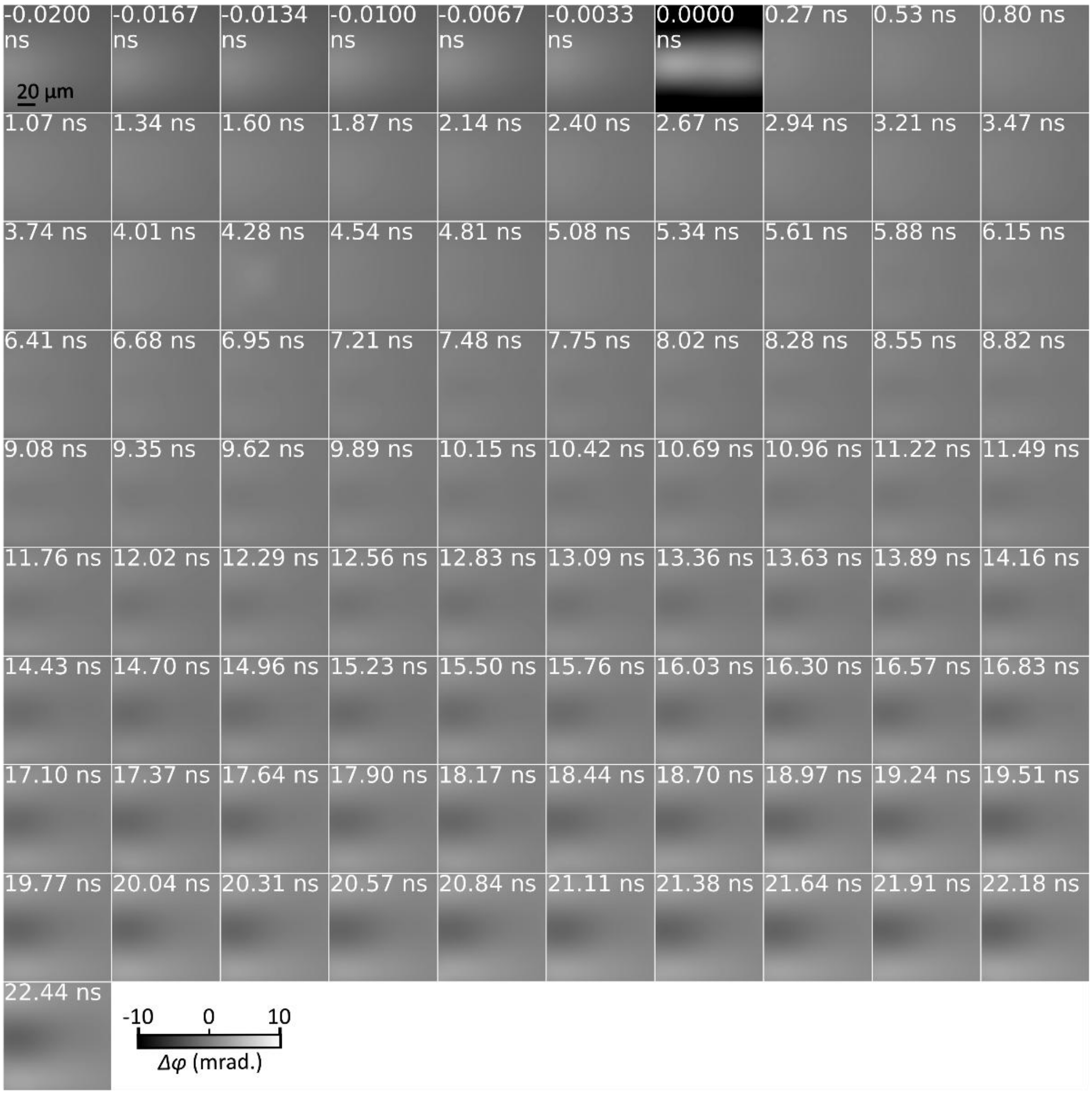

**Supplementary Figure 4,** Background fit of the differential phase images for the water experiments discussed in Figure 4 of the manuscript.

**Supplementary Information 3, Determining wave-propagation speeds,**

Light-induced waves can be either classified as pressure- or shock-waves. The former exhibit speeds of sound as governed by the material properties, the latter speeds that exceed the nominal speed of sound. To identify the nature of our waves (Figure 4), we estimate its time-dependent radius using a Hough transformation as shown in Supplementary Figure 5. By fitting the radius as a function of time we obtain a wave propagation velocity of approximately 1438 m/s, closely resembling the speed of sound in water.

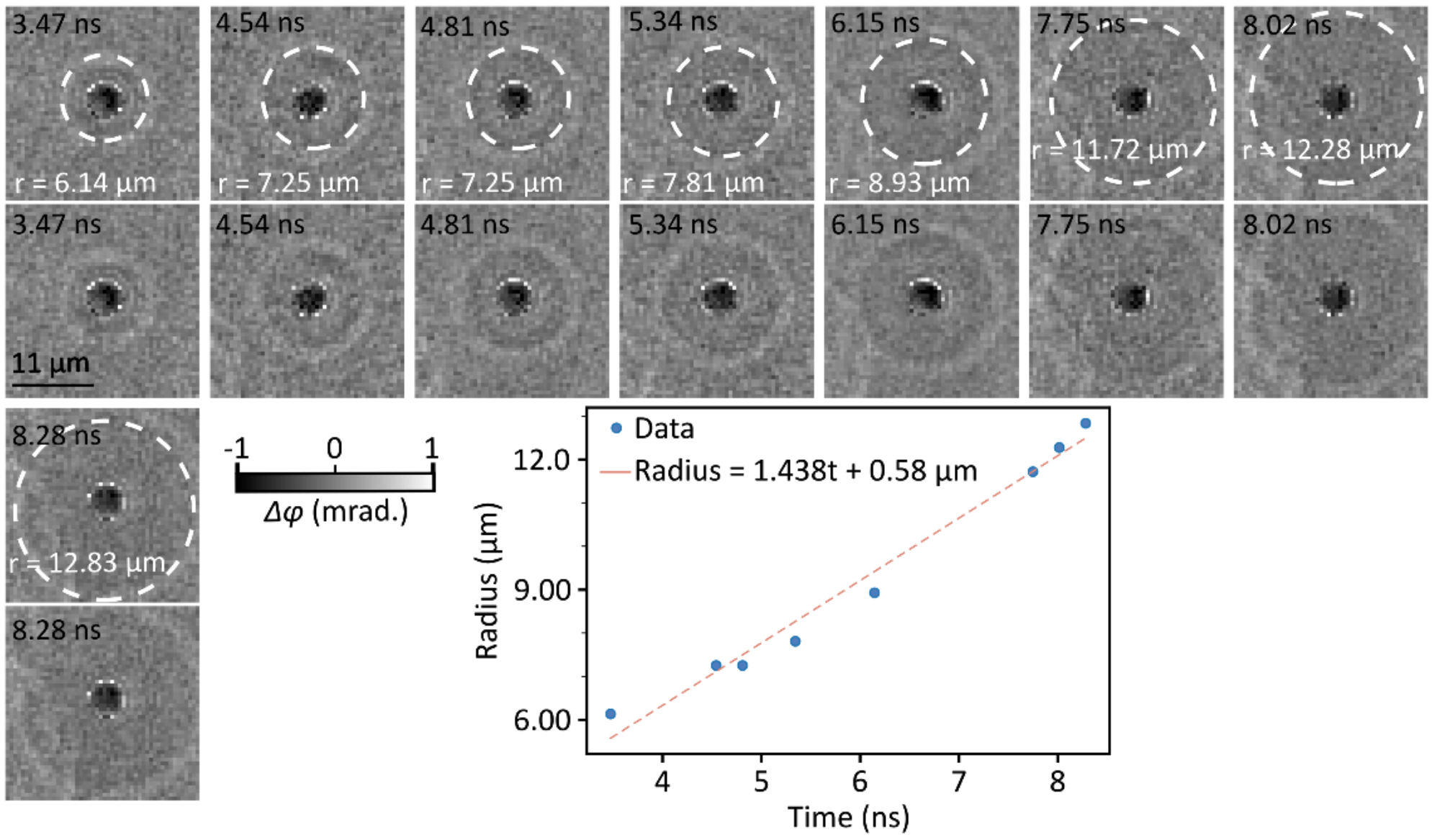

**Supplementary Figure 5.** Radius of the induced pressure wave determined via Hough transformations as a function of time, together with a linear fit to estimate the speed of sound.

**Supplementary Information 4, processing oscillatory data for zero-padding and re-fitting 5 µm sphere oscillations,**

As the FFT's shown in Figure 3 in the manuscript provide only rough centre frequencies due to its coarse FFT-binning, we process the oscillatory data by removing a slight polynomial baseline trend and applying a gaussian Hanning window, allowing for zero-padding with minimal artifacts (Supplementary Figure 6a). In Figure 6c in the manuscript we fit our oscillatory data using a sum of two simple damped oscillators fixing the predicted frequencies in Figure 6b, using the nominal diameters of the PS spheres, where a slight dephasing can be seen at the end of the oscillatory trace for the 5 µm sphere. Changing the diameter of the sphere to 5.14 µm we obtain the predicted oscillation as seen in Supplementary Figure 6b, removing this minor dephasing and identifying the slight diameter difference as a likely explanation.

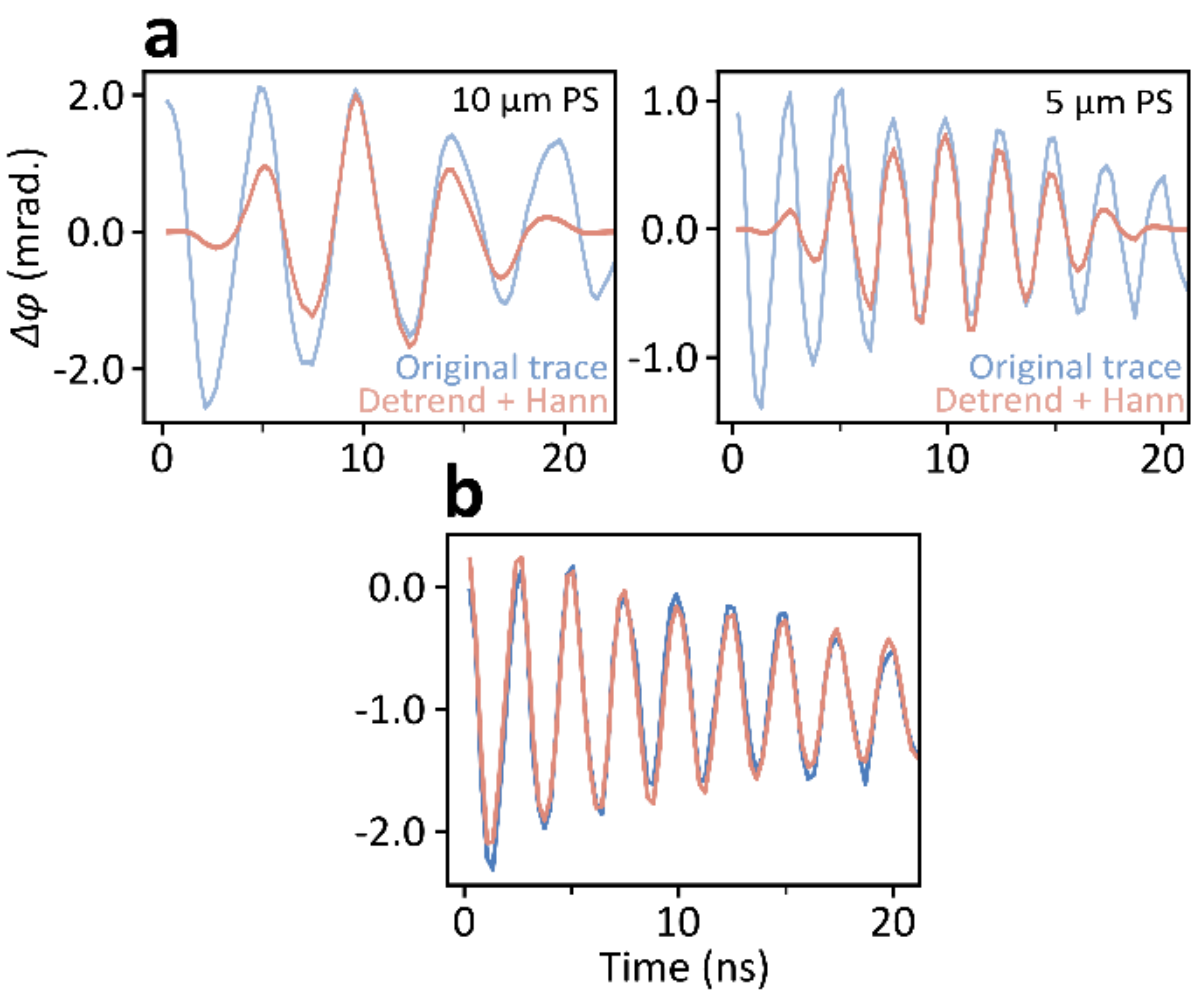

**Supplementary Figure 6.** a) Original vs processed oscillatory data, the latter allowing for zero-padding with minimal artifacts. b) Measured oscillatory data of the 5 µm sphere fit with the sum of two damped oscillators fixed with the predicted frequencies at a diameter of 5.14 µm.